\documentclass[conference]{IEEEtran}
\IEEEoverridecommandlockouts
\usepackage{cite}
\usepackage{amsmath,amssymb,amsfonts}
\usepackage{algorithmic}
\usepackage{graphicx}
\usepackage{textcomp}
\usepackage{xcolor}

\usepackage{amsmath}
\usepackage{graphicx}
\usepackage{booktabs}
\usepackage{array}
\usepackage{url}
\usepackage{tikz}
\usetikzlibrary{shapes.geometric, arrows.meta, positioning, fit, calc}
\usepackage{textcase}
\usepackage[colorlinks=true, linkcolor=blue, citecolor=blue,
            urlcolor=blue]{hyperref}

\providecommand{\ntag}[1]{\texttt{#1}}

\def\BibTeX{{\rm B\kern-.05em{\sc i\kern-.025em b}\kern-.08em
    T\kern-.1667em\lower.7ex\hbox{E}\kern-.125emX}}
\begin{document}

\title{VAANI Noise Event Dataset: A curated spontaneous speech dataset annotated with timestamps for noise events\\
{\footnotesize \textsuperscript{*}}
\thanks{}
}

% \and
% \and
% \and
% \and
% \and

\author{
\IEEEauthorblockN{
Pavan Kumar J\textsuperscript{1},
Agneedh Basu\textsuperscript{1},
Pranav Bhat\textsuperscript{1},
Sujith Pulikodan\textsuperscript{1},
Suryansh Shukla\textsuperscript{1},
Nihar Desai\textsuperscript{1},\\
Prasanta K. Ghosh\textsuperscript{2}
}
\IEEEauthorblockA{
\textsuperscript{1}AI \& Robotics Technology Park (ARTPARK), I-Hub @ IISc, Bangalore, India\\
\textsuperscript{2}Department of Electrical Engineering, Indian Institute of Science, Bangalore, India
}
}

\maketitle

\begin{abstract}
Most public sound-event corpora are optimised either for general audio
tagging or for clean speech separation, and comparatively few provide
strong (timestamped) noise annotations layered directly on top of
spontaneous, real-world speech. We present the \emph{VAANI Noise Event
Timestamp Dataset} , a derived annotation
layer built on Project VAANI field recordings of
spontaneous speech collected across 165 Indian districts in 105
languages. Unlike synthetically mixed corpora, VAANI
captures speech and ambient noise \emph{in~situ} and simultaneously, and
annotates each recording with fine-grained start/end timestamps for
overlapping background noise events organised into a compact seven-class
semantic taxonomy (animal, traffic, baby/child, music, signal/alarm,
appliance, and non-speech human). This combination---spontaneous
multilingual Indic speech, authentic regional soundscapes, and
span-level noise tags that may overlap with speech---targets tasks that
existing datasets address only partially: noise-robust Automatic Speech
Recognition (ASR), sound event detection (SED), and speech enhancement.
We position VAANI against nine widely used corpora and benchmarks
(WHAM!, AVA-Speech, MUSAN, FSD50K, CHiME-6, AudioSet, DESED, the
India-specific iNoise noise database, and the Kathbath-Noisy noisy-ASR
benchmarks, among others) and describe the annotation protocol and
quality-control procedure used to produce
the timestamped tags.
\end{abstract}


% =====================================================================
\section{Introduction}
% =====================================================================
Automatic speech recognition (ASR) and related speech technologies are
increasingly deployed in everyday Indian settings, yet the audio they
encounter there rarely resembles the clean, studio-quality material on
which many models are trained. Speech is captured on commodity mobile
devices in household kitchens, on farms, on busy local streets, and in
crowded indoor spaces, where it co-occurs with a rich and highly
non-stationary background: vehicle horns and passing traffic, animal
calls, crying infants, appliances, music, alarms, and a variety of human
non-speech sounds. This background is not merely additive noise
energy---its onset, duration, and overlap with speech directly shape
recognition errors and the perceived quality of enhancement. Building
systems that are robust to such conditions therefore requires data that
reflect them faithfully, capturing not only the \emph{acoustic content}
of the noise but also when each noise event occurs relative to
the spoken signal.

A number of influential corpora each capture part of this picture, but
along different axes and with different objectives: synthetically
mixed corpora such as WHAM!~\cite{wham} and DESED's~\cite{desed}
synthetic subset; frame- or clip-level general-audio corpora such as
AVA-Speech~\cite{avaspeech}, FSD50K~\cite{fsd50k}, and
AudioSet~\cite{audioset}; isolated sub-corpora such as
MUSAN~\cite{musan}; spontaneous-but-domain-mismatched corpora such as
CHiME-6~\cite{chime6}; and India-specific resources such as the iNoise
noise database~\cite{inoise} and the Kathbath-Noisy noisy-ASR
benchmarks~\cite{kathbath-noisy}. Section~\ref{sec:related} reviews each
of these in detail and contrasts them with VAANI; in short, none
combines real, in-situ co-occurrence of speech and background noise,
span-level overlapping event timestamps, and spontaneous multilingual
speech collected under the single-channel, mobile-device conditions
typical of large-scale Indian field data collection---the combination
few resources provide.

The VAANI Noise Event Timestamp Dataset is designed to fill this gap. It
is a derived annotation layer on top of Project VAANI's~\cite{vaani}
spontaneous speech recordings, adding exact start/end timestamps for background
noise events---for example \ntag{horn} or \ntag{dog\_bark}---alongside the
speech transcript, orienting the dataset toward noise-robust ASR, sound
event detection, and speech enhancement in realistic Indian acoustic
conditions.

% =====================================================================
\section{Related Datasets}
\label{sec:related}
% =====================================================================
Table~\ref{tab:compare} places VAANI alongside nine widely used noise,
sound-event, and noisy-ASR resources, contrasting how each was recorded,
what kind of speech (if any) it contains, how noise is annotated, and
its scale. Two terms recur throughout the table: \emph{in-situ} audio is
recorded naturally in the field, with speech and background noise
co-occurring together as they would in real life, rather than being
synthetically mixed after separate recording; and \emph{span-level}
annotation gives each noise event an exact start and end timestamp,
rather than only a clip-level or frame-level label. These nine resources
fall into four groups by construction method, discussed in turn below.

\subsection{Synthetically mixed corpora}
WHAM!~\cite{wham} takes the WSJ0-2mix speech-separation benchmark and
adds a synthetic ambient-noise channel. WSJ0-2mix itself comprises
20{,}000, 5{,}000, and 3{,}000 instantaneous two-speaker mixtures in its
30-, 10-, and 5-hour train, validation, and test sets respectively
(45 hours of clean, read WSJ0 speech in total); training and validation
share speakers, while the test speakers are disjoint from both. WHAM!
mixes this speech with roughly 80 hours of urban background audio,
independently recorded with binaural microphones at San Francisco Bay
Area cafes, bars, and offices, algorithmically combined at controlled
SNRs to produce 28{,}000 two-speaker noisy mixtures. Because the two
channels are recorded separately and combined after the fact, WHAM!
provides exact waveform-level ground truth for both speaker and
background signals, but the pairing is synthetic by construction and
carries no event-level taxonomy of what kind of noise is present.
DESED~\cite{desed} is a
hybrid: 17{,}850 real 10-second clips drawn from AudioSet
($\sim$49.6 hours split across weakly labelled, unlabelled in-domain,
validation, and public-evaluation subsets) are supplemented with
additional synthetic soundscapes generated by the Scaper tool from a
bank of 2{,}060 background and 1{,}009 foreground recordings, targeting
10 domestic sound-event classes with weak, strong, or unlabelled
annotation depending on the subset. Neither corpus offers real,
naturally co-occurring foreground speech alongside its noise
annotations---WHAM!'s speech is clean and read, and DESED's clips are
domestic soundscapes without an annotated speech signal at all---which
is the pairing VAANI is built to capture.

\subsection{Frame- and clip-level general-audio corpora}
AVA-Speech~\cite{avaspeech} labels roughly 45 hours of movie audio at
the frame level with four mutually exclusive speech-activity
states---clean speech, speech+music, speech+noise, and no
speech---aggregated across three human raters with substantial
inter-annotator agreement. This scheme flags \emph{that} noise co-occurs
with speech in a given frame but not \emph{which} category of noise is
present or its precise onset and offset. FSD50K~\cite{fsd50k} and
AudioSet~\cite{audioset} take the opposite tack: both provide only weak,
clip-level multi-label tags---200 AudioSet-ontology classes over
51{,}197 Freesound clips (108 hours) for FSD50K, and 527 classes over
more than two million 10-second YouTube clips for AudioSet---with no
timestamp information and, particularly for AudioSet, little control
over recording conditions, language, or the presence of speech at all.
All three corpora sacrifice the event-level timing precision that
VAANI's \{category, tag, start, end\} annotation format is designed to
preserve.

\begin{table*}
  \centering
  \caption{VAANI versus representative noise / sound-event corpora.}
  \label{tab:compare}
  \small
  \setlength{\tabcolsep}{4pt}
  \begin{tabular}{@{}p{2.1cm}p{3.0cm}p{2.6cm}p{4.0cm}p{3.2cm}@{}}
    \toprule
    \textbf{Corpus} & \textbf{Audio origin} & \textbf{Speech /style} &
      \textbf{Noise annotation} & \textbf{Scale} \\
    \midrule
    WHAM!~\cite{wham} & Synthetic mix: clean read speech (WSJ0) + urban
      ambient noise & Read, single-speaker & Waveform-level source /
      background ground truth, no event taxonomy &
      28{,}000 two-speaker mixtures; $\sim$80\,h noise \\
    AVA-Speech~\cite{avaspeech} & Movie soundtracks & dialogue only &
      Frame-level speech-activity state (clean / +music / +noise / none),
      4 mutually exclusive classes & $\sim$45\,h \\
    MUSAN~\cite{musan} & Isolated speech, music, and noise sub-corpora &
      Read / none & None---speech and noise are separate, non-overlapping
      subsets & $\sim$109\,h \\
    FSD50K~\cite{fsd50k} & Freesound web clips &
      Incidental (voice/speech among the 200 classes) &
      Weak, clip-level labels, 200 AudioSet-ontology classes &
      51{,}197 clips; 108\,h \\
    AudioSet~\cite{audioset} & YouTube clips &
      Incidental (`Speech' a frequent class) &
      Weak, clip-level labels, 527 classes &
      2M+ 10\,s clips \\
    DESED~\cite{desed} & Domestic soundscapes, recorded (AudioSet) +
      Scaper-synthesized & Incidental (`speech' = 1 of 10 classes) &
      Weak / strong / unlabelled subsets,
      10 domestic SED classes &
      $\sim$49.6\,h: weak 4.4\,h,
      unlabeled 40.0\,h, val 3.2\,h, eval 1.9\,h + synthetic clips  \\
    CHiME-6~\cite{chime6} & Real dinner parties, multi-channel
      Kinect/binaural arrays & Spontaneous, overlapping, multi-speaker
      English & None (diarization / ASR target, not noise-labelled) &
      40+\,h, 20 sessions \\
    iNoise~\cite{inoise} & Indian indoor/outdoor environmental noise
      recordings & None (noise-only, no speech) &
      Noise-type category per file only; no co-occurring speech, no
      event timestamps &
      10 categories (5 outdoor, 5 indoor); 1 concatenated file/category \\
    Kathbath-Noisy~\cite{kathbath-noisy} & Read speech (AI4Bharat
      Kathbath) + injected noise, per-language ASR benchmark sets &
      Read, single-speaker, per Indian language & None---ASR
      transcription (word-level) only, no noise category/timestamp
      labels &
      Kathbath base: 1{,}684\,h across 12 Indian languages; per-language
      noisy test sets (e.g.\ Odia, Tamil, Bengali) \\
    \midrule
    \textbf{VAANI Noise Event Timestamp Dataset} & \textbf{In-situ field recording, single
      channel, mobile devices} & \textbf{Spontaneous, multilingual
      Indic} & \textbf{Strong, span-level, overlapping timestamps; 7
      ASR-oriented classes} & \textbf{72{,}756 segs, 122.17\,h, 58 langs,
      30 states} \\
    \bottomrule
  \end{tabular}
\end{table*}

\begin{figure*}
  \centering
  \resizebox{\textwidth}{!}{%
  \begin{tikzpicture}[
      block/.style={rectangle, draw, thick, fill=blue!6, rounded corners,
        text width=2.5cm, text centered, minimum height=1.1cm,
        font=\scriptsize, inner sep=3pt},
      decision/.style={diamond, draw, thick, fill=orange!12, aspect=1.7,
        text width=2.1cm, text centered, font=\scriptsize, inner sep=1pt},
      outcome/.style={rectangle, draw, thick, fill=green!10,
        rounded corners, text width=2.4cm, text centered,
        minimum height=1.1cm, font=\scriptsize, inner sep=3pt},
      lbl/.style={font=\tiny},
      arr/.style={thick, -{Latex[length=2mm]}},
      node distance=9mm and 9mm
    ]
    \node[block] (corpus) {1. VAANI corpus\\($\sim$150\,h+ sampled)};
    \node[block, right=of corpus] (freelance)
      {2. Freelancer\\timestamps};
    \node[decision, right=of freelance] (sanity)
      {3. Sanity check:\\pass?};
    \node[outcome, right=of sanity] (relU)
      {Release $\sim$100\,h \textbf{unverified} };

    \node[block, below=of sanity] (predata)
      {$\sim$20\,h+ pre-data for verified pool};
    \node[block, right=of predata] (qc)
      {4. Internal QC\\timestamp};
    \node[decision, right=of qc] (audit)
      {5. 10\% random\\independent QC:\\pass?};
    \node[outcome, right=of audit] (relV)
      {Release \textbf{verified}};

    \draw[arr] (corpus) -- (freelance);
    \draw[arr] (freelance) -- (sanity);
    \draw[arr] (sanity) -- node[above, lbl]{pass} (relU);
    \draw[arr] (sanity) -- node[right, lbl]{pass} (predata);
    \draw[arr] (predata) -- (qc);
    \draw[arr] (qc) -- (audit);
    \draw[arr] (audit) -- node[above, lbl]{pass} (relV);

    \draw[arr] (sanity.north) -- ++(0cm,0.7cm)
      -| node[pos=0.22, above, lbl]{fail: redo batch (\textrightarrow\,2)}
      (freelance.north);
    \draw[arr] (audit.south) -- ++(0cm,-0.7cm)
      -| node[pos=0.22, below, lbl]{fail: redo QC (\textrightarrow\,4)}
      (qc.south);
  \end{tikzpicture}}
  \caption{Annotation and QC pipeline: freelancer output that passes a
  sanity check splits into a $\sim$100-hour \texttt{unverified} 
  release and a $\ge$20-hour subset that is internally re-timestamped
  and 10\%-audited before release as \texttt{verified\_timestamps}
  ; each failure loops back to the preceding step.}
  \label{fig:qc-pipeline}
\end{figure*}
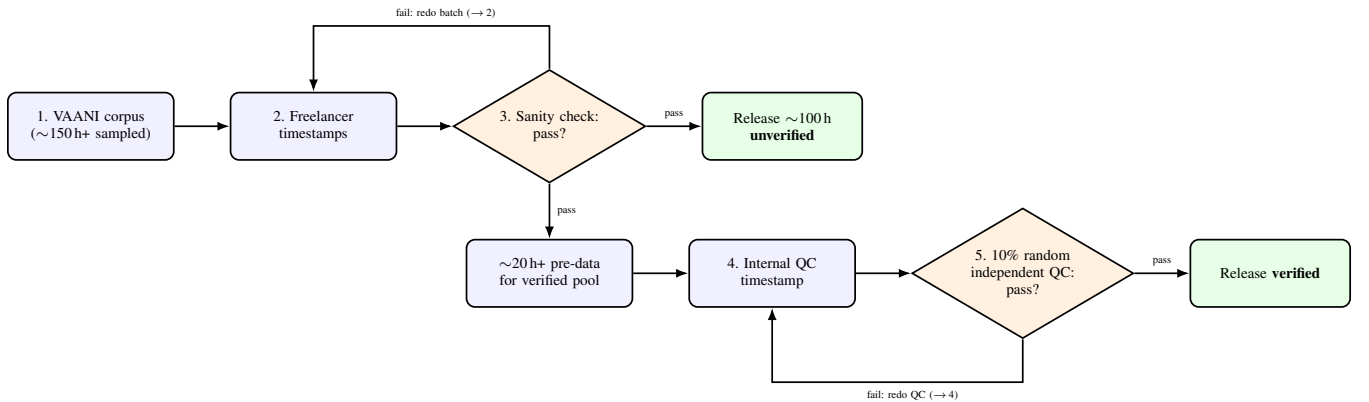

\subsection{Isolated and domain-mismatched spontaneous-speech corpora}
MUSAN~\cite{musan} assembles about 109 hours of speech, music, and noise
into three \emph{separate} sub-corpora built for voice-activity
detection and music/speech discrimination; because speech and noise
never co-occur in the same recording, MUSAN cannot support the joint
speech-plus-noise-event task VAANI targets. CHiME-6~\cite{chime6} is the
closest existing corpus to VAANI in spirit---more than 40 hours of
unscripted, overlapping, multi-speaker conversational speech recorded
across 20 real dinner parties using multi-channel Kinect arrays and
binaural microphones---but its target is diarization and distant
multi-speaker ASR in English-language Western domestic settings, and it
carries no noise-event category or timestamp annotations at all.

\subsection{Indian noise and noisy-ASR resources}
Two India-specific resources address the domain mismatch identified
above but not the annotation gap. iNoise~\cite{inoise} is a database of
ten Indian environmental noise-type categories---five outdoor
(autorickshaw, bus, highway, railway station, street) and five indoor
(airport, cafeteria, home, train, workplace)---recorded specifically
because environmental noise in Indian conditions differs substantially
from the noise typically studied in Western corpora. It is, however, a
noise-only recording set: it contains no co-occurring speech and no
annotation of when a noise event occurs relative to an utterance, so it
is typically used to synthetically corrupt clean speech for robustness
testing rather than to study natural speech-noise co-occurrence. The
Kathbath-Noisy family of benchmarks~\cite{kathbath-noisy} (e.g.\
Kathbath-Odia-Noisy), released on IndiaAI's AIKosh platform and built on
AI4Bharat's Kathbath corpus (1{,}684 hours of read speech across 12
Indian languages), evaluates ASR systems on noisy speech per language.
These benchmarks pair noisy audio with ground-truth transcriptions for
measuring word-error-rate degradation under noise, but---like
CHiME-6---their labels are speech transcriptions, not noise-event
categories or timestamps; what kind of noise is present, and exactly
when, is not annotated.

\subsection{Positioning}
Across all nine corpora and benchmarks, no single resource combines
(i) real, in-situ co-occurrence of speech and background noise,
(ii) span-level, overlapping timestamps for the noise events themselves,
and (iii) spontaneous, multilingual speech collected under the
single-channel, mobile-device conditions typical of large-scale Indian
field data collection. Even the two India-specific resources above
satisfy the geography and domain criteria but not the annotation
criterion: iNoise has no speech at all, and Kathbath-Noisy labels only
the transcription, not the noise event. VAANI is positioned to close
this specific combination of gaps, as summarised in the rightmost column
of Table~\ref{tab:compare}.

\begin{table*}
  \begin{minipage}[t]{0.47\textwidth}
    \centering
    \parbox[t][7em][t]{\linewidth}{\centering
    \caption{Dataset overview (release-quality subset: \texttt{hasIssue}
    IS NULL, \texttt{NoiseCategory} IS NOT NULL, \texttt{evalSet} and
    \texttt{syntheticData} both NOT TRUE, \texttt{annotationQuality}
    $\in$ \{verified\_timestamps, unverified\_timestamps\}).}
    \label{tab:overview}}
    \small
    \begin{tabular}{@{}lr@{}}
      \toprule
      \textbf{Property} & \textbf{Value} \\
      \midrule
      Segments & 72{,}756 \\
      Total audio & 122.17 h \\
      Distinct speakers & 38{,}541 \\
      Languages & 58 \\
      States / districts & 30 / 162 \\
      Segment duration (min/mean/max) & 0.79 / 6.05 / 23.49 s \\
      Segments with a noise category & 72{,}756 \\
      Segments with timestamped events & 72{,}746 \\
      Total timestamped noise events & 106{,}892 \\
      \addlinespace
      \quad Verified timestamps   & 11{,}111 segs / 21.85 h \\
      \quad Unverified timestamps  & 61{,}645 segs / 100.32 h \\
      \bottomrule
    \end{tabular}
  \end{minipage}
  \hfill
  \begin{minipage}[t]{0.47\textwidth}
    \centering
    \parbox[t][7em][t]{\linewidth}{\centering
    \caption{Noise categories: segment coverage and event statistics
    over the same 72{,}756-segment subset as Table~\ref{tab:overview}
    (categories may co-occur, so ``\% seg.'' need not sum to 100\%).}
    \label{tab:categories}}
    \scriptsize
    \setlength{\tabcolsep}{3.5pt}
    \begin{tabular}{@{}lrrrr@{}}
      \toprule
      \textbf{Category} & \textbf{Seg.} & \textbf{\% seg.} &
        \textbf{Events} & \textbf{Ev. dur.} \\
      \midrule
      Non-speech human   & 27{,}533 & 37.8 & 37{,}739 & 4.5 h \\
      Animal             & 22{,}746 & 31.3 & 24{,}601 & 22.4 h \\
      Vehicle / traffic  & 18{,}150 & 24.9 & 20{,}603 & 12.5 h \\
      Baby / child       & 11{,}708 & 16.1 & 12{,}376 & 10.5 h \\
      Singing / music    &  6{,}879 &  9.5 &  6{,}978 &  9.7 h \\
      Phone / signal / alarm & 3{,}546 & 4.9 & 3{,}683 & 1.8 h \\
      Appliance / machine &   906 & 1.3 &   912 & 1.5 h \\
      \bottomrule
    \end{tabular}
  \end{minipage}
\end{table*}

% =====================================================================
\section{Dataset Description}
\label{sec:dataset}
% =====================================================================
The dataset comprises \textbf{72{,}756 speech segments} totalling
\textbf{122.17 hours} of audio, drawn from \textbf{38{,}541 distinct
speakers}. This is the release-quality subset: segments with a
structural issue flag, held-out eval-set segments, synthetic-data
segments are excluded, leaving only the
\texttt{verified\_timestamps}  and \texttt{unverified\_timestamps}
 tiers. Segments are short utterances---0.79 to 23.49 seconds
long, with a mean of 6.05 seconds---each carrying a speech transcript
and one or more timestamped background-noise events. Coverage is
geographically and linguistically broad: \textbf{58 languages} across
\textbf{30 states} and \textbf{162 districts} of India, spanning both
collection phases of Project VAANI. Table~\ref{tab:overview} summarises
these headline figures.

\subsection{Linguistic and geographic coverage}
The corpus is dominated by Hindi (83.9 h; 47{,}080 segments) but retains a
substantial long tail: Telugu (16.7 h), Bengali (12.9 h), Marathi
(5.3 h), Nepali, Malayalam, Assamese, Kannada, Odia, and lower-resource
languages such as Chakma, Garo, and Mizo, among the 57 in total. At the
regional level, Bihar (24.0 h), Andhra Pradesh (14.3 h), Uttar Pradesh
(12.9 h), West Bengal (12.3 h), and Maharashtra (11.9 h) lead, with the
remaining hours spread across 25 further states. This spread ensures that
the background soundscapes are drawn from a wide variety of rural,
semi-urban, and urban Indian acoustic environments rather than a single
locale.

\subsection{Noise taxonomy and annotation format}
Each segment is annotated at two levels. The \emph{segment level} records
the set of noise categories present as a multi-label list
(\texttt{NoiseCategory}); the \emph{event level} records, for each noise
occurrence, a \{category, tag, start, end\} tuple with exact start/end
timestamps in seconds (\texttt{NoiseSubCategoryTimeStamp}). Timestamps
are stored as verbatim-precision strings, and the underlying tag (e.g.
\ntag{barking}, \texttt{[lip smacking]}) is preserved alongside its
canonical category. Events may overlap one another and co-occur with
speech.

Seven top-level noise categories are represented.
Table~\ref{tab:categories} reports, for each, the number of segments in
which it appears and the number and total duration of its timestamped
events. \emph{Non-speech human} sounds are the most frequent by segment
coverage (37.8\% of all segments) and by event count (37{,}739 events),
but---being brief interjections such as coughs and lip smacks---account
for only 4.5 event-hours at a mean of 0.42\,s per event. \emph{Animal}
and \emph{vehicle/traffic} noises, by contrast, are longer-lived
(mean 3.3\,s and 2.2\,s) and together contribute the bulk of annotated
noise duration. \emph{Appliance/machine} events are the rarest but the
longest on average (6.1\,s).

All \textbf{72{,}756} segments carry at least one noise category (by
construction of the release-quality filter) and \textbf{72{,}746} carry
at least one timestamped event, yielding \textbf{106{,}892} noise events
in total. Most segments contain a single noise category (\textbf{55{,}330}),
but a substantial minority exhibit two or more co-occurring categories
(\textbf{17{,}426}), reflecting the layered soundscapes of real field
recordings.

% \subsection{Annotation tiers}
% Segments carry an \texttt{annotationQuality} tier reflecting the depth of
% human verification. \textbf{Verified timestamps}  9{,}256 segments,
% 16.3 h) have human-checked event boundaries; \textbf{unverified
% timestamps} (; 63{,}120 segments, 102.6 h) carry timestamped events
% that have not undergone the full  review; and \textbf{no timestamps}
% (17{,}963 segments, 32.6 h) provide category labels without per-event
% spans. A held-out evaluation split of 2{,}129 segments is flagged for
% benchmarking, and 968 segments are marked with a structural
% \texttt{hasIssue} flag for downstream filtering.

% =====================================================================
\section{Annotation and Quality Control}
\label{sec:qc}
% =====================================================================
Noise-event timestamps are produced by an external freelancer pool and
pass through a staged internal quality-control (QC) pipeline before
release. Figure~\ref{fig:qc-pipeline} summarises the pipeline; the
numbered steps below correspond to the diagram.

\textbf{1. Sampling.} Roughly 150+ hours of segments are sampled from
the broader VAANI spontaneous-speech corpus for noise-event annotation.

\textbf{2. Freelancer timestamping.} A pool of trained freelancers
listens to each sampled segment and marks the start/end timestamp and
category tag for every audible noise event.

\textbf{3. Sanity check.} The complete freelancer output is checked for
structural validity. If it fails, the entire batch is sent back to step
2 and redone. A batch that passes is split: about 100 hours are released
directly as \texttt{unverified\_timestamps} , and a further
$\ge$20-hour subset is carried forward as candidate data for the
verified tier.

\textbf{4. Internal QC timestamping.} An internal team with the reference of \texttt{unverified\_timestamps} 
re-timestamps the $\ge$20-hour candidate subset.

\textbf{5. Independent 10\% audit.} A second, independent reviewer
audits a random 10\% sample of the internally re-timestamped subset. If
even a single event in that sample disagrees, the batch is sent back to
step 4 and redone; only once the audit passes is the subset released as
\texttt{verified\_timestamps}.

This staged design---a full sanity gate before any release, plus an
independently re-timestamped and 10\%-audited candidate subset for the
gold tier, with each failure looping back to the immediately preceding
step---is why \texttt{verified\_timestamps} carries a materially higher
trust level than \texttt{unverified\_timestamps}, even though both tiers
consist of fully timestamped events (Table~\ref{tab:overview}).

% % =====================================================================
% \section{Target Tasks and Use Cases}
% \label{sec:tasks}
% % =====================================================================
% \emph{[Placeholder: noise-robust ASR, sound event detection, and speech
% enhancement use cases.]}

% =====================================================================
\section{Conclusion}
\label{sec:conclusion}
% =====================================================================
We introduced the VAANI Noise Event Timestamp Dataset, a span-level
noise-annotation layer over Project VAANI's in-situ, spontaneous
multilingual Indic speech recordings. Unlike synthetically mixed
corpora (WHAM!, DESED's synthetic subset) or corpora with only
frame-/clip-level noise labels (AVA-Speech, FSD50K, AudioSet), VAANI
pairs naturally co-occurring speech and background noise with exact,
overlapping start/end timestamps organised into a compact seven-class,
ASR-oriented taxonomy (Table~\ref{tab:compare}). The release-quality
subset comprises 72{,}756 segments (122.17 hours, 38{,}541 speakers)
across 58 languages, 30 states, and 162 districts of India
(Table~\ref{tab:overview}), with non-speech human, animal, and
vehicle/traffic events dominating segment coverage and animal and
vehicle/traffic events contributing the bulk of annotated noise
duration (Table~\ref{tab:categories}).

Every timestamp passes through the staged annotation and QC pipeline
described in Section~\ref{sec:qc} (Figure~\ref{fig:qc-pipeline}): a
sanity-checked freelancer pool produces the \texttt{unverified\_timestamps} tier, and a further internally re-timestamped, 10\%-audited
subset is released as the higher-trust \texttt{verified\_timestamps}
 tier, giving downstream users an explicit, auditable quality
distinction rather than a single undifferentiated label set. By
combining this scale and geographic/linguistic breadth with strong,
overlapping noise annotations on authentic field speech, VAANI is
positioned to support noise-robust ASR, sound event detection, and
speech enhancement research on the Indian acoustic conditions that
existing public corpora leave largely uncovered.

% \vspace{12pt}
% \color{red}
% IEEE conference templates contain guidance text for composing and formatting conference papers. Please ensure that all template text is removed from your conference paper prior to submission to the conference. Failure to remove the template text from your paper may result in your paper not being published.

\end{document}